\documentclass[prl,twocolumn,superscriptaddress]{revtex4}
\usepackage{epsfig}
\usepackage{times}
\usepackage{amsmath}
\usepackage{amsfonts}
\usepackage{amssymb}
\usepackage[usenames]{color}
\usepackage{graphicx}
\newcommand{\NTT}{NTT Basic Research Laboratories, NTT Corporation, 3-1 Morinosato-Wakamiya, Atsugi, Kanagawa, 243-0198, Japan.}

\newcommand{\keio}{Department of Applied Physics and Department of Physico-Informatics, Faculty of Science and Technology,
Keio University, Hiyoshi, Kohoku-ku, Yokohama 223-8522, Japan}
\newcommand{\beq}{\begin{equation}}
\newcommand{\eeq}{\end{equation}}
\newcommand{\beqa}{\begin{eqnarray}}
\newcommand{\eeqa}{\end{eqnarray}}

\renewcommand{\vec}[1]{\ensuremath{\mathbf{#1}}}

\begin{document}
\title{
Vector magnetic field sensing via multi-frequency control of
nitrogen-vacancy centers in diamond
}
\author{Sayaka Kitazawa}
   \affiliation{
\keio
   }
\author{Yuichiro Matsuzaki}
   \affiliation{
\NTT
   }
 \author{Saijo Soya
 }
 \affiliation{\keio}
 \author{Kosuke Kakuyanagi}
   \affiliation{
\NTT
   }
    \author{Shiro Saito}
   \affiliation{
\NTT
   }
 \author{Junko Ishi-Hayase
 }
 \affiliation{\keio}

\begin{abstract}
An ensemble of nitrogen-vacancy (NV) centers in diamond is an attractive
 %technique
 \textcolor{black}{device}
 to detect small magnetic fields.
 %Since the NV centers have
 %four different orientations
In particular, by exploiting the fact that the NV center can be aligned along one of
four different axes due to $C_{3\nu } $ symmetry, it is possible to
 extract information concerning vector magnetic
fields. However, in the conventional scheme, low readout
 contrasts of the NV centers significantly decrease the sensitivity of
 the vector magnetic field sensing.
 Here, we propose a way to improve the sensitivity of the vector
 magnetic field sensing of the NV centers using multi-frequency
 control.
 %Because
 \textcolor{black}{Since}
 the Zeeman energy of the NV centers depends on the direction of the axis, we can
 independently control the four types of NV centers using microwave
 pulses with different frequencies. This allows us to use every NV center
 for the vector field detection in parallel, which effectively
 increases the readout contrast. Our results pave the way to realize a
 practical diamond-based vector field sensor.
\end{abstract}

\maketitle

%\section{I. introduction}

%Measurement of small magnetic fields is an important objective in the field of
%metrology because of many practical applications in material science, biology, and
%medical science.
 The detection of small magnetic fields is \textcolor{black}{important} in the
field of metrology, because there are many potential applications in
biology and medical science.
The performance of a magnetic field sensor is characterized by its spatial
resolution and sensitivity; therefore, a significant amount of effort has been
devoted to creating a device that can measure small magnetic fields in a
local region \cite{simon1999local,chang1992scanning,poggio2010force}.
%A two-level system  can be used for the
%magnetic fields sensing if the frequency of the two-level system is
%shifted by the magnetic fields.

Nitrogen vacancy (NV) centers in diamond are fascinating candidates with which to
construct a magnetic field sensor \cite{maze2008nanoscale, taylor2008high, balasubramanian2008nanoscale, schaffry2011proposed}. The NV center is a spin 1 system, and
the frequency of the $|\pm 1\rangle$ states can be
shifted by magnetic fields.
We can use this system as an effective two-level system spanned by
$|0\rangle $ and $|1\rangle$ with a frequency selectivity where
$|-1\rangle $ is significantly detuned.
 We can implement gate operations of the spins in NV
 centers using microwave pulses 
 \cite{Go01a,gruber1997scanning,jelezko2002single,JGPGW01a}.
It is possible to detect DC (AC)
magnetic fields by implementing a Ramsey interference (spin echo)
measurement \cite{maze2008nanoscale, taylor2008high, balasubramanian2008nanoscale}.
Moreover, NV centers have a long coherence time, e.g., a few milli-seconds
at a room temperature and a second at low temperature
\cite{balasubramanian2009ultralong,mizuochi2009coherence,bar2013solid}. In addition,
because the NV centers can be strongly coupled with optical photons, we can read out the state of the NV centers via
fluorescence
from the optical transitions  \cite{gruber1997scanning,jelezko2002single}. The NV centers can be embedded in
nanocrystals, which allows the NV centers to interact with local magnetic
fields \cite{schirhagl2014nitrogen}.
These properties are prerequisite
to realizing a high-performance sensor for magnetic fields.

Recently, vector magnetic field sensing by NV centers has become an active area of
interest \cite{maertz2010vector,steinert2010high,pham2011magnetic,tetienne2012magnetic,dmitriev2016concept,sasaki2016broadband}.
%we can use NV centers to sense vector magnetic fields.
The NV center is aligned along one of
four different axes due to $C_{3\nu }$ symmetry.
%By applying external
%known magnetic
%field,
The Zeeman energies of the NV centers are determined by $g\mu
_b\vec{B}\cdot \vec{d}_j$ $(j=1,2,3,4)$ where $g$ denotes the g factor,
$\mu _b$ denotes a Bohr magneton, $\vec{B}$ denotes  the magnetic
fields, and $\vec{d}_j$ denotes the direction of the $j$-th NV axis.
By sequentially performing Ramsey interference or spin echo
measurements on NV centers with different NV axes, we
can estimate the values of the Zeeman energies $g\mu
_b\vec{B}\cdot \vec{d}_j$. The data
from the experiments can be processed to reconstruct
the vector components ($B_x$, $B_y$, and $B_z$) of applied magnetic
fields \cite{maertz2010vector,pham2011magnetic}.
This can be used to magnetically image a target sample such as living
cells or circuit currents \cite{le2013optical,nowodzinski2015nitrogen}.

%However,
In the conventional approach, the low readout
 contrast of the NV centers decreases the sensitivity when sensing
 the vector magnetic field \cite{taylor2008high,acosta2009diamonds}.
 When the state of the NV centers is $|\pm 1\rangle $, the photoluminescence
 intensity becomes smaller than in the case of $|0\rangle$.
 This allows us to measure the state of the NV centers via optical
 detection even at room temperature.
 % To construct an ideal projective
%  measurement, the NV centers should emit the photons if and only if the
%  state is $|0\rangle$, which generates a correlation between the photon
%  detection and the state of $|0\rangle$. Nevertheless, at room
%  temperature, the state of $|\pm 1\rangle $ also emits photons.
% Although we can measure the state of
 %the NV centers by fluorescence
%from the optical transitions,
Nevertheless, we can only detect a small portion of the
 emitted photons, because most of the photons are emitted into the
 environment. %unless we confine the NV centers into a high-Q cavity.
 This decreases the readout contrast. Moreover, if we only implement
 Ramsey or spin echo measurements on NV centers with a specific axis
 with this limited readout contrast,
 the states of the other NV centers with different axes remain in the
 $|0\rangle$ state regardless the value of the magnetic fields, which
 induces noise affecting the sensitivity of the magnetic field
 sensor \cite{taylor2008high}.
 If we only need to estimate one vector component of the target magnetic
 field, we can recover the sensitivity by using a diamond where the
 orientations of the NV centers are aligned along just one axis \cite{michl2014perfect,lesik2014perfect,fukui2014perfect}.
However, we cannot use such a diamond to estimate every component of
 the vector magnetic fields, unless we mechanically rotate the diamond to
 change the angle between the target magnetic fields and the direction
 of the NV axis.
  \begin{figure}[h] 
\begin{center}
\includegraphics[width=0.99\linewidth]{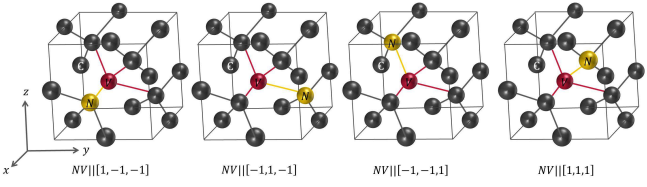} 
\caption{An NV center in diamond set an axis to be the
direction from the vacancy to the nitrogen. There are
 four possible directions of the axis in the diamond.
 Note that, by applying known external magnetic
 fields, we can independently control the four types of NV centers with
 different axes using frequency selectivity.
 }
 \label{schematic}
\end{center}
\end{figure}

 Here, we propose a scheme to improve the sensitivity of the vector
 magnetic field sensing
via multi-frequency control. Because NV centers with different axes can
 have different resonant frequencies \cite{acosta2009diamonds}, we can independently control these NV centers via frequency selectivity. 
The key idea in our scheme is the simultaneous implementation of a Ramsey interference or
 spin echo experiment with every NV center by via multi- frequency
 control.
 We show that adequate control of the microwave pulses can
 enhance the signal from NV centers with four different axes, and that the
 sensitivity of the vector magnetic field sensing becomes approximately four times better
 than that of the conventional scheme.

 \section{Conventional vector magnetic field sensing with an NV center}
 %\subsection{DC magnetic field sensing}

   \begin{figure}[h] 
\begin{center}
\includegraphics[width=0.99\linewidth]{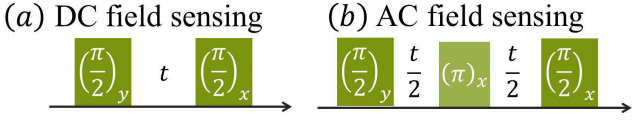} 
\caption{Microwave pulse sequence for standard magnetometry with NV
 centers; 
(a) Ramsey interference measurements performed to sense DC field,
 and (b) Spin echo measurements performed to sense the AC field where we
 can suppress low
 frequency magnetic field noise.
 }
 \label{previouspulse}
\end{center}
\end{figure}
 Here, we review conventional DC magnetic field sensing using NV
 centers \cite{taylor2008high,maertz2010vector,steinert2010high}.
 Even though the NV center is a spin-1 system, we can treat it as a
 two-level system spanned by $|0\rangle $ and $|1\rangle $ with 
 frequency selectivity.
   Note that the NV center has four types of intrinsic quantization axes along
  the NV direction with zero or small magnetic fields. We define the
  direction of these NV axes as
 $\vec{d}_1=(\frac{1}{\sqrt{3}},-\frac{1}{\sqrt{3}},-\frac{1}{\sqrt{3}})$,
 $\vec{d}_2=(-\frac{1}{\sqrt{3}},\frac{1}{\sqrt{3}},-\frac{1}{\sqrt{3}})$,
  $\vec{d}_3=(-\frac{1}{\sqrt{3}},-\frac{1}{\sqrt{3}},\frac{1}{\sqrt{3}})$, and
    $\vec{d}_4=(\frac{1}{\sqrt{3}},\frac{1}{\sqrt{3}},\frac{1}{\sqrt{3}})$. 
  %$\vec{d}_k$ $(k=1,2,3,4)$.
 The Hamiltonian of the NV center with an axis defined by a vector $\vec{d}_k$ is given as
  \begin{eqnarray}
   H_k=\frac{\omega _k}{2}\hat{\sigma }_z+\lambda  \hat{\sigma }_x\cos
    \omega _k't\label{hamiltonian}
  \end{eqnarray}
  where $\omega _k=\omega _0+g\mu _b \vec{B}_{\rm{total}}\cdot \vec{d}_k$, $\omega _0$
  denotes a zero field splitting,  $g\mu _b
  \vec{B}_{\rm{total}}\cdot \vec{d}_k$ denotes a Zeeman energy
  splitting,
  %$g$ denotes a g-factor, $\mu _b$ denotes a Bohr magneton,
  $\vec{B}_{\rm{total}}=\vec{B}_{\rm{ex}}+\vec{B}$ denotes the sum of
  a known external magnetic field ($\vec{B}_{\rm{ex}}$) and the target
  unknown magnetic
  field ($\vec{B}$), $\lambda $
  denotes a Rabi frequency, and $\omega '_k$ denotes a microwave frequency.
  In a rotating frame, we can rewrite this Hamiltonian as
  \begin{eqnarray}
   H_k=\frac{\omega_k -\omega _k'}{2}\hat{\sigma }_z+\frac{\lambda}{2}  \hat{\sigma }_x\label{free}
  \end{eqnarray}
  where we choose $\omega '_k=\omega _0 +g\mu _b \vec{B}_{\rm{ex}}\cdot
  \vec{d}_k$.
  If we do not apply a microwave, the Hamiltonian in the rotating frame
  is written as
  \begin{eqnarray}
   H_k=\frac{g\mu _b\vec{B}\cdot \vec{d}_k}{2}\hat{\sigma }_z
  \end{eqnarray}
  We can construct the vector field sensor as follows (see Fig. \ref{previouspulse}(a)). We assume
  that the initialization time, pulse operations, and readout time are
  much shorter than the coherence time of the NV center.
  First, we initialize $|0\rangle_k$state via green laser irradiation.
  Second, by performing a $\frac{\pi }{2}$ pulse along the $y$ axis with the microwave, we
  prepare a $|+\rangle _k=\frac{1}{\sqrt{2}}(|0\rangle_k
  +|1\rangle _k)$ state.
  Third, we let this state evolve via the Hamiltonian in 
  Eq. (\ref{free}) for a time $t_k$. Note that the NV center is affected
  by the dephasing process; therefore, the dynamics can be described by the
  following master equation.
  \begin{eqnarray}
   \frac{d\rho _k}{dt}=-i[H_k,\rho _k]-\gamma _k(\rho _k-\hat{\sigma }_z\rho_k
    \hat{\sigma }_z)\label{master}
  \end{eqnarray}
  where $\gamma _k=\frac{1}{2T_{2}^{k*}}$ denotes the dephasing rate of the $k$ the NV center and
  $T_{2}^{k*}$ denotes the coherence time  measured by
  Ramsey interference.
%   By solving this master equation, we obtain
%   \begin{eqnarray}
%    &&\rho _k(t)=\frac{1}{2}|1\rangle \langle 1|+\frac{1}{2}|0\rangle
%     \langle 1|e^{-2\gamma t+ig\mu _b\vec{B}\cdot \vec{d}_k t}\nonumber \\
%     &+&\frac{1}{2}|1\rangle \langle 0|e^{-2\gamma t-ig\mu _b\vec{B}\cdot \vec{d}_k t}+\frac{1}{2}|0\rangle \langle 0|
%   \end{eqnarray}
Fourth, we perform a $\frac{\pi }{2}$ pulse along the $x$ axis with the
microwave.
The diagonal component of the density matrix after these operations can be calculated to be
\begin{eqnarray}
{}_k \langle 0|\rho _k(t_k)|0\rangle _k=\frac{1+e^{-2\gamma _kt_k}\sin (g\mu
  _b\vec{B}\cdot \vec{d}_kt_k)}{2}\label{diagonalzero} \\
  {}_k\langle 1|\rho _k(t_k)|1\rangle _k=\frac{1-e^{-2\gamma _kt_k}\sin (g\mu
  _b\vec{B}\cdot \vec{d}_kt_k)}{2}. \label{diagonalone}
\end{eqnarray}
Finally, we readout the population of the state via the green laser
irradiation \cite{gruber1997scanning}.
The information of the NV center is now transferred into photons, and
 the photon state is described as
\begin{eqnarray}
 \rho ^{(\rm{ph})}_k&=&\frac{1+e^{-2\gamma _kt_k}\sin (g\mu
  _b\vec{B}\cdot \vec{d}_kt_k)}{2} \rho ^{(\rm{ph})}_{k,0}\nonumber \\
  &+&\frac{1-e^{-2\gamma _kt_k}\sin (g\mu
  _b\vec{B}\cdot \vec{d}_kt_k)}{2} \rho ^{(\rm{ph})}_{k,1}
\end{eqnarray}
where $\rho ^{(\rm{ph})}_{k,1}$ ($\rho ^{(\rm{ph})}_{k,0}$) denotes the
 state of the photon after performing the green laser pulse when the
state of the NV center is $|0\rangle _k$ ($|1\rangle _k$).
We can describe $\rho ^{(\rm{ph})}_{k,1}$ and $\rho ^{(\rm{ph})}_{k,0}$
as follows.
\begin{eqnarray}
  \rho ^{(\rm{ph})}_{k,0}&=&(1-\tilde{\alpha }^{(k)}_0)|0\rangle
  _{\rm{ph}}\langle 0|
  +\tilde{\alpha }^{(k)}_0|1\rangle
  _{\rm{ph}}\langle 1|\nonumber \\
  \rho ^{(\rm{ph})}_{k,1}&=&(1-\tilde{\alpha }^{(k)}_1)|0\rangle
  _{\rm{ph}}\langle 0|+\tilde{\alpha }^{(k)}_1|1\rangle
  _{\rm{ph}}\langle 1| \nonumber
\end{eqnarray}
where $|0\rangle_{\rm{ph}}$ and $|1\rangle_{\rm{ph}}$ denote the Fock states
of the photon. We define
\begin{eqnarray}
 \tilde{\alpha }^{(k)}_0&=&\sum_{j=1}^{4}\alpha ^{(j)}_0 \nonumber \\
  \tilde{\alpha }^{(k)}_1&=&\alpha ^{(k)}_1+\sum_{j\neq k}\alpha ^{(j)}_0 \nonumber \\
\end{eqnarray}
where $\alpha ^{(k)}_0$ ($\alpha ^{(k)}_1$) for $k=1,2,3,4$ denotes the
probability emitting a photon when the state of the NV center is
$|0\rangle _k$ ($|1\rangle_k$). Note that, while we
control the NV center with the NV axis along $\vec{d}_k$, the other NV
centers remain the $|0\rangle $ state and emit photons.
 We assume
$\alpha ^{(k)}_0,\ \alpha ^{(k)}_1 \ll 1$ and that the multiple photon
emission probability from an NV center is negligible.
For $g\mu
   _b\vec{B}\cdot \vec{d}_kt_k\ll 1$, we can calculate the expectation value of the emitted photons
\begin{eqnarray}
 \langle \hat{N}_k\rangle&=&{\rm{Tr}}[\rho ^{(\rm{ph})}_k
  \hat{N}]\nonumber \\
%  &=&\frac{1-e^{-2\gamma _kt_k}\sin (g\mu
%    _b\vec{B}\cdot \vec{d}_kt_k)}{2}\tilde{\alpha }_0^{(k)}\nonumber \\
%   &+&\frac{1+e^{-2\gamma _kt_k}\sin (g\mu
%    _b\vec{B}\cdot \vec{d}_kt_k)}{2}\tilde{\alpha }_1^{(k)}\nonumber \\
  &\simeq &\frac{1+ g\mu
   _b\vec{B}\cdot \vec{d}_kt_ke^{-2\gamma _kt_k}}{2}\tilde{\alpha }_0^{(k)}\nonumber \\
  &+&\frac{1- g\mu
   _b\vec{B}\cdot \vec{d}_kt_ke^{-2\gamma _kt_k}}{2}\tilde{\alpha }_1^{(k)}
\end{eqnarray}
where $\hat{N}=\sum_{n=0}^{\infty }|n\rangle _{\rm{ph}}\langle n|$.
Note that we can tune $\alpha ^{(k)}_0$ and $\alpha
^{(k)}_1$ by changing both the applied known magnetic fields and
the polarization of the photons. In addition, we can decrease the coherence time if we add
artificial noise.
For simplicity, we assume $\alpha ^{(k)}_0=\alpha _0$, $\alpha
^{(k)}_1=\alpha _1$, $\gamma _k=\gamma $, and $t_k=t$ for all
$k$. Suppose we
first implement the above experiment shown in 
Fig. \ref{previouspulse}(a) for $k=1$, and then implement
it for $k=4$, which allows us to sum up these two experimental data.
We
obtain
\begin{eqnarray}
&&\langle \hat{N}_1\rangle+\langle
 \hat{N}_4\rangle\nonumber \\
%  &=&\frac{2-\sum_{k=1,4} e^{-2\gamma t} (g\mu
%    _b\vec{B}\cdot \vec{d}_kt)}{2}\tilde{\alpha }_0\nonumber \\
%   &+&\frac{2+\sum_{k=1,4} e^{-2\gamma t} (g\mu
%    _b\vec{B}\cdot \vec{d}_kt)}{2}\tilde{\alpha }_1\nonumber \\
%   &=&\frac{2- e^{-2\gamma t} \frac{2}{\sqrt{3}}g\mu
%    _bB_xt}{2}\tilde{\alpha }_0
%   +\frac{2+ e^{-2\gamma t} \frac{2}{\sqrt{3}}g\mu
%    _bB_xt}{2}\tilde{\alpha }_1\nonumber \\
 &=&(\tilde{\alpha }_0+\tilde{\alpha }_1)+\frac{1}{\sqrt{3}}
  (\tilde{\alpha }_0-\tilde{\alpha }_1)g\mu _bB_xte^{-2\gamma t}
\end{eqnarray}
Interestingly, this sum depends on $B_x$
 \textcolor{black}{while this} is independent of $B_y$ and $B_z$.
Therefore we define $\langle \hat{N}_x\rangle
 \equiv \langle \hat{N}_1\rangle+\langle
 \hat{N}_4\rangle$, and we estimate $B_x$ from $\langle \hat{N}_x\rangle$.
Note that, even though we explain the case to measure
$B_x$, we can also measure $B_y$ ($B_z$) by
considering $\langle \hat{N}_y\rangle
 \equiv \langle \hat{N}_2\rangle+\langle
 \hat{N}_4\rangle $ ($\langle \hat{N}_z\rangle
 \equiv \langle \hat{N}_3\rangle+\langle
 \hat{N}_4\rangle $), because $\langle \hat{N}_y\rangle$ ($\langle
 \hat{N}_z\rangle$) only depends on $B_y$ ($B_z$).
Therefore, we can calculate the uncertainty in the estimation of $B_x$
as follows.
\begin{eqnarray}
 \delta B^{\rm{(DC)}}_x&=&\frac{\sqrt{\langle \delta \hat{N}_x  \delta
  \hat{N}_x\rangle }}{|\frac{d\langle \hat{N}\rangle_x
  }{dB_x}|}\frac{1}{\sqrt{N}}\nonumber \\
%  &=&\frac{\sqrt{\tilde{\alpha }_0+\tilde{\alpha }_1
%   }}{\frac{1}{\sqrt{3}} e^{-2\gamma
%   t}g\mu _bt|\tilde{\alpha }_0-\tilde{\alpha
%   }_1|}\frac{1}{\sqrt{\frac{T}{2t}}}\nonumber \\ 
  &=&\frac{\sqrt{3}\sqrt{7\alpha _0+\alpha _1
  }}{ |\alpha _0-\alpha _1|
  g\mu _bte^{-2\gamma t}}\frac{1}{\sqrt{\frac{T}{2t}}}
\end{eqnarray}
where $N=\frac{T}{2t}$ denotes the repetition number and $T$ denotes the 
total experiment time. This uncertainty is minimized for
$t=\frac{1}{4\gamma }$ and, 
\begin{eqnarray}
 \delta B^{\rm{(DC)}}_x
  &=&\frac{\sqrt{3}\sqrt{7\alpha _0+\alpha _1
  }}{ e^{-\frac{1}{4}}|\alpha _0-\alpha _1|
  g\mu _b\sqrt{\frac{1}{4\gamma }}}\frac{1}{\sqrt{\frac{T}{2}}}\label{conventional}
\end{eqnarray}
Therefore we chose this value for the field
sensing.
Note that we have a factor of
$\sqrt{7\alpha _0+\alpha _1}$ in the numerator, which increases the
uncertainty. This is because,
 when we readout the NV centers, three quarters of the NV
centers remain in the $|0\rangle$ state regardless of strength of
the magnetic fields, which decreases the sensitivity.
This clearly shows that the existence of NV centers that emit the same amount of 
photons regardless of the strength of the applied magnetic field actually
decreases the sensitivity of the field sensing.
% It is worth mentioning that, although we explained how to measure a
% magnetic field $B_x$ along $[1,0,0]$, it is easy to generalize this
% scheme to measure $B_y$ and $B_z$.
% To measure $B_y$, we need to 

Here, we briefly review conventional AC magnetic field sensing
using NV centers \cite{taylor2008high,maertz2010vector,steinert2010high}.
We have the same form of the Hamiltonian described in Eq. \ref{hamiltonian}
%The Hamiltonian is described as follows.
% \begin{eqnarray}
%    H_k&=&\frac{\tilde{\omega }_k}{2}\hat{\sigma }_z+\lambda  \hat{\sigma }_x\cos
%     \tilde{\omega} _k't\nonumber \\
%   \end{eqnarray}
  where
  % $\tilde{\omega} _k=\omega _0+g\mu _b \tilde{\vec{B}}_{\rm{total}}\cdot \vec{d}_k$, 
  % $g\mu _b
%   \vec{B}_{\rm{total}}\cdot \vec{d}_k$ denotes a Zeeman energy
%   splitting,
we replace the total magnetic field with
  $\vec{B}_{\rm{total}}=\vec{B}_{\rm{ex}}+\vec{B}_{\rm{AC}}\sin
  \omega _{\rm{AC}}t$.
%   denotes a sum of
%   a known external magnetic field ($\vec{B}_{\rm{ex}}$) and target
%   unknown magnetic
%   field ($\vec{B}_{\rm{AC}}$) with a frequency of $\omega
%   _k^{\rm{(AC)}}$.
  % ,
%   and 
%   $\tilde{\omega} '_k=\omega _0 +g\mu _b \vec{B}_{\rm{ex}}\cdot
%   \vec{d}_k$ denotes a microwave frequency.
To estimate the values of $\vec{B}_{\rm{AC}}$,
  we use a similar pulse sequence to that of the DC magnetic field sensing.
The only difference from the DC magnetic field sensing is that we apply
a $\pi$ pulse in the middle of the time evolution between the two
$\frac{\pi}{2}$ pulse, as shown in Fig. \ref{previouspulse} (b).
The diagonal component of the density matrix can be calculated as
\begin{eqnarray}
&&{}_k \langle 0|\rho _k(t_k)|0\rangle _k
 =\frac{1+e^{-2\gamma '_kt_k}\sin \theta _k^{\rm{(AC)}}}{2}\label{diagonalzeroac} \\
  &&{}_k\langle 1|\rho _k(t_k)|1\rangle _k=\frac{1-e^{-2\gamma '_kt_k}\sin
   \theta _k^{\rm{(AC)}}}{2}. \label{diagonaloneac}\\
 &&\theta _k^{\rm{(AC)}}=
  g\mu_b\vec{B}\cdot \vec{d}_k\frac{1+\cos \omega _{\rm{AC}}t-2\cos \frac{\omega _{\rm{AC}}t}{2}}{\omega _{\rm{AC}}}
\end{eqnarray}
where $\gamma _k'=\frac{1}{2T^k_{2}}$ denotes the dephasing rate for the
$k'$ th NV center and $T^k_{2}$ denotes the dephasing time measured by the
spin echo.
% The states of the photon after the
% green laser irradiation are described as
% \begin{eqnarray}
%  &&\rho _k^{\rm{(ph)}}=\Big{(}
%   (1-\alpha _0^{(k)})\frac{1-e^{-2\gamma '_kt_k}\sin \theta
%   _k^{\rm{(AC)}}}{2}\nonumber \\
%   &+&
%   (1-\alpha _1^{(k)})\frac{1+e^{-2\gamma '_kt_k}\sin \theta _k^{\rm{(AC)}}}{2}
%   \Big{)}
%   |0\rangle _{\rm{ph}}\langle 0|\nonumber \\
%  &+&\Big{(}
%   \alpha _0^{(k)}\frac{1-e^{-2\gamma '_kt_k}\sin \theta
%   _k^{\rm{(AC)}}}{2}\nonumber \\
%  &+&
%   \alpha _1^{(k)}\frac{1+e^{-2\gamma '_kt_k}\sin \theta _k^{\rm{(AC)}}}{2}
%   \Big{)}
%   |1\rangle _{\rm{ph}}\langle 1|
% \end{eqnarray}
Similar to the case of DC sensing, we can calculate the sensitivity of the AC
field sensing such that
\begin{eqnarray}
 \delta B^{\rm{(AC)}}_x
  \simeq \frac{\sqrt{3}\sqrt{7\alpha _0+\alpha _1
  }}{ |\alpha _0-\alpha
  _1|g\mu _b\frac{|1+\cos \omega _{\rm{AC}}t-2\cos \frac{\omega
  _{\rm{AC}}t}{2}|}{\omega _{\rm{AC}}}e^{-2\gamma'
  t}}\frac{1}{\sqrt{\frac{T}{2t}}}\ \ 
\end{eqnarray}
where we assume $\alpha ^{(k)}_0=\alpha _0$, $\alpha
^{(k)}_1=\alpha _1$, $\gamma '_k=\gamma '$, and $t_k=t$ for all $k$.
This uncertainty is minimized for
$t=\frac{1}{4\gamma '}$ and $\omega _{\rm{AC}}\simeq 23.3\gamma
'=\frac{2\theta _{\rm{opt}}}{T_2}$ for $\theta _{\rm{opt}}\simeq 1.856
\pi $,
%($\omega _{\rm{AC}}t=5.83$),
;therefore, we choose these values for the field sensing.
The uncertainty in the estimation is given as follows.
\begin{eqnarray}
 \delta B^{\rm{(AC)}}_x
  \simeq \frac{\sqrt{3}\sqrt{7\alpha _0+\alpha _1
  }}{ e^{-\frac{1}{2}}|\alpha _0-\alpha
  _1|g\mu _b\frac{|1+\cos \theta _{\rm{opt}}-2\cos \frac{\theta
  _{\rm{opt}}}{2}|}{\theta _{\rm{opt}}}\sqrt{\frac{1}{4\gamma '}}}\frac{1}{\sqrt{\frac{T}{2}}}\label{conventionalac}\ \ 
\end{eqnarray}

\section{DC vector magnetic field sensor via multi-frequency control}
Here, we propose a scheme to measure the vector magnetic field with
an improved sensitivity. The key idea is to adopt multi-frequency
control of the NV centers. NV centers with different axes can
 have different resonant frequencies when applying a known external magnetic
 field \cite{acosta2009diamonds}; therefore, we can independently control
 these NV centers using frequency
 selectivity.
 In addition, we can parallelize the control of the NV centers
 by simultaneously rotating all NV centers with different axes so
 that every NV center can be involved in the field sensing.
% We will show that the combination of these ideas actually makes the
% sensitivity approximately 4 times better than the conventional scheme.
% The key idea in our scheme is to implement the Ramsey interference of
%  spin echo experiment for all NV centers with different NV axes in
%  parallel.
%  We have shown that an adequate control of the microwave pulses can
%  enhance the signal from NV center with four different axes, and the
%  sensitivity of the vector magnetic field sensing becomes 4 times better
%  than that of the conventional scheme.

  \begin{figure}[h] 
\begin{center}
\includegraphics[width=1.03\linewidth]{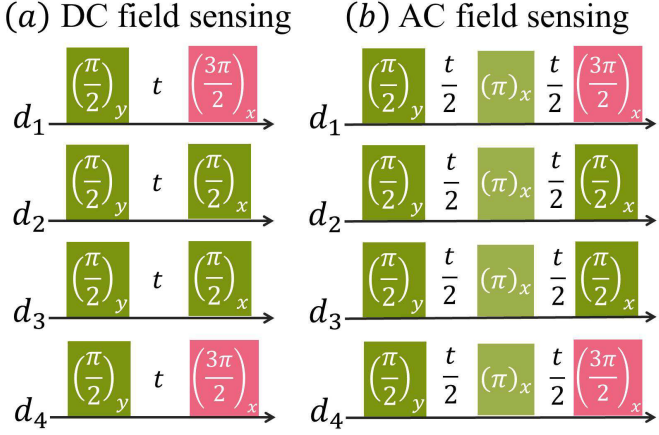} 
\caption{The pulse sequence used to perform our proposed vector magnetic field
 sensing. Using frequency selectivity, we independently control the NV
 centers with different axes. We implement four microwave pulses with different
 frequencies at the same time to increase the sensitivity.
 }
 \label{ourpulse}
\end{center}
\end{figure}

As an example, we explain how to measure a DC magnetic field component along $[1,0,0]$ ($B_x$) using our scheme.
%In our scheme,
After the initialization of the states by the green
laser, we rotate every NV center using the $\frac{\pi }{2}$ pulse, and
the initial state is given by
\begin{eqnarray}
 \bigotimes _{k=1}^4\frac{1}{\sqrt{2}}(|0\rangle _k+|1\rangle _k).
\end{eqnarray}
We let this state evolve for a time $t$ according to the master equation in 
Eq. (\ref{master}).
After performing the $\frac{\pi }{2}$ pulse ($\frac{3\pi }{2}$ pulse) on the NV centers with the NV
axes $\vec{d}_2$ and $\vec{d}_3$ ($\vec{d}_1$ and $\vec{d}_4$) as
shown in Fig. \ref{ourpulse}(a), we
read out the
state of the NV centers via the photoluminescence.
The diagonal component of the density matrix just before the readout can be calculated to be
\begin{eqnarray}
{}_k \langle 0|\rho _k(t_k)|0\rangle _k=\frac{1+e^{-2\gamma _kt_k}\sin (g\mu
  _b\vec{B}\cdot \vec{d}_kt_k)}{2}\nonumber \\
  {}_k\langle 1|\rho _k(t_k)|1\rangle _k=\frac{1-e^{-2\gamma _kt_k}\sin (g\mu
  _b\vec{B}\cdot \vec{d}_kt_k)}{2}
\end{eqnarray}
for $k=2,3$ and
\begin{eqnarray}
{}_{k} \langle 0|\rho _{k}(t_{k})|0\rangle _{k}=\frac{1-e^{-2\gamma _{k}t_{k}}\sin (g\mu
  _b\vec{B}\cdot \vec{d}_{k}t_{k})}{2}\nonumber \\
  {}_{k'}\langle 1|\rho _{k}(t_{k})|1\rangle _{k}=\frac{1+e^{-2\gamma _{k}t_{k}}\sin (g\mu
  _b\vec{B}\cdot \vec{d}_{k}t_{k})}{2}
\end{eqnarray}
for $k=1,4$.
After the green laser irradiation, the state of the photons can be
described as follows.
\begin{eqnarray}
 \rho ^{\rm{(ph)}}= \bigotimes _{k=1}^4\rho _k^{\rm{(ph)}}
\end{eqnarray}
where
\begin{eqnarray}
 \rho ^{(\rm{ph})}_k&=&\frac{1+e^{-2\gamma _kt_k}\sin (g\mu
  _b\vec{B}\cdot \vec{d}_kt_k)}{2} \rho ^{(\rm{ph})}_{k,0}\nonumber \\
  &+&\frac{1-e^{-2\gamma _kt_k}\sin (g\mu
  _b\vec{B}\cdot \vec{d}_kt_k)}{2} \rho ^{(\rm{ph})}_{k,1}
\end{eqnarray}
for $k=2,3$ and
\begin{eqnarray}
 \rho ^{(\rm{ph})}_{k}&=&\frac{1-e^{-2\gamma _{k}t_{k}}\sin (g\mu
  _b\vec{B}\cdot \vec{d}_{k}t_{k})}{2} \rho ^{(\rm{ph})}_{k,0}\nonumber \\
  &+&\frac{1+e^{-2\gamma _{k}t_{k}}\sin (g\mu
  _b\vec{B}\cdot \vec{d}_{k}t_{k})}{2} \rho ^{(\rm{ph})}_{k,1}
\end{eqnarray}
for $k=1,4$.
We can calculate the expected values of the emitted photons from these
states as follows:
\begin{eqnarray}
 \langle\hat{N}_x^{\rm{(total)}}\rangle
  &=&{\rm{Tr}}[(\sum_{k=1}^{4}\hat{N}_k)\rho ^{\rm{(ph)}}]\nonumber \\
 &\simeq &
    \sum_{k=1}^{4}\frac{\alpha _0^{(k)}+\alpha
  _1^{(k)}}{2}\nonumber \\
  &-&\sum_{k=1,4}\frac{\alpha _0^{(k)}-\alpha _1^{(k)}}{2} g\mu
   _b\vec{B}\cdot \vec{d}_{k}t_{k}e^{-2\gamma _{k}t_{k}}\nonumber \\
  &+&\sum_{k=2,3}\frac{\alpha _0^{(k)}-\alpha _1^{(k)}}{2} g\mu
   _b\vec{B}\cdot \vec{d}_{k}t_{k}e^{-2\gamma _{k}t_{k}}\ \ \ \ 
   \label{ninhomodc}
\end{eqnarray}
If we have $\alpha ^{(k)}_0=\alpha _0$, $\alpha
^{(k)}_1=\alpha _1$, $\gamma _k=\gamma $, and $t_k=t$ for all $k$, we obtain
\begin{eqnarray}
\langle\hat{N}_x^{\rm{(total)}}\rangle  \simeq 2(\alpha _0+\alpha
 _1)-
 \frac{2}{\sqrt{3}}(\alpha _0-\alpha _1)g\mu _bB_xte^{-2\gamma t}
  \nonumber
\end{eqnarray}
Note that this expectation values depends on just $B_x$.
%where $\hat{N}_k= \sum_{n=0}^{\infty }|n\rangle _{{\rm{ph}},k}\langle
%n|$
Therefore, the uncertainty of the estimation of $B_x$ is given as
follows.
\begin{eqnarray}
 \delta B^{(\rm{DC})}_x&=& \frac{\sqrt{\langle \delta \hat{N}_x^{\rm{(total)}}  \delta
  \hat{N}_x^{\rm{(total)}}\rangle }}{|\frac{d\langle \hat{N}_x^{\rm{(total)}}\rangle
  }{dB_x}|}\frac{1}{\sqrt{N}}\nonumber \\
 &\simeq &\frac{\sqrt{2(\alpha _0+\alpha _1)}}{\frac{2}{\sqrt{3}}|\alpha _0-\alpha _1|g\mu _bte^{-2\gamma t}}\frac{1}{\sqrt{\frac{T}{t}}}\label{our}
\end{eqnarray}
where $N=\frac{T}{t}$ denotes the repetition number of the experiment.
This uncertainty is minimized for
$t=\frac{1}{4\gamma }$ and
\begin{eqnarray}
 \delta B^{(\rm{DC})}_x
 &\simeq &\frac{\sqrt{3}\sqrt{2(\alpha _0+\alpha
 _1)}}{2e^{-\frac{1}{2}}|\alpha _0-\alpha _1|g\mu
 _b\sqrt{\frac{1}{4\gamma }}}\frac{1}{\sqrt{T}}\label{our}
\end{eqnarray}
Therefore, we chose this value for the field
sensing.
Because we have $\alpha _0\simeq \alpha _1$ due to the low readout
contrast \cite{acosta2009diamonds}, the
sensitivity of our scheme described by (Eq. \ref{our}) is approximately four times better than that in the
conventional scheme described by
Eq. (\ref{conventional}).
Note that, even though we explained how to measure the 
magnetic field $B_x$ along $[1,0,0]$, we can easily generalize our
scheme to measure $B_y$ and $B_z$. For example, to measure $B_y$ ($B_z$),
we perform a $\frac{\pi }{2}$ pulse ($\frac{3\pi }{2}$ pulse) on the NV centers with the NV
axes of $\vec{d}_1$ and $\vec{d}_3$ ($\vec{d}_2$ and $\vec{d}_4$)
between the green laser irradiation.

  \begin{figure}[h] 
\begin{center}
\includegraphics[width=0.9\linewidth]{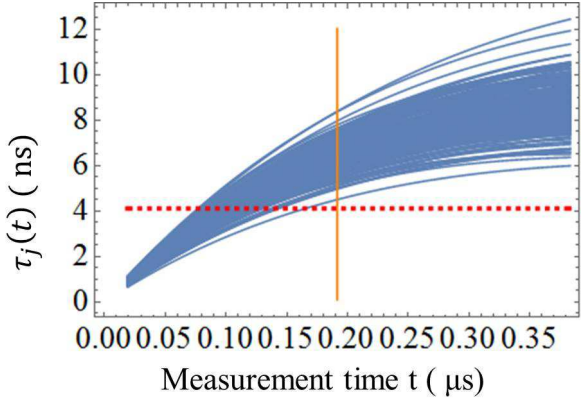} 
\caption{We plot $\tau _j(t)=\frac{\alpha _0^{(j)}-\alpha
_1^{(j)}}{2}e^{-2\gamma _{j}t}t$ ($j=1,2,\cdots ,200$) against $t$ where
 we choose $\delta \alpha _j=\alpha
_0^{(j)}-\alpha _1^{(j)}$ and $\gamma _j$ from the Gaussian
distribution. The average of $\delta \alpha _j$ ($\gamma _j$) is $0.01$
 ($10^6$ Hz), and the
 standard deviation is $0.001$ ($10^5$).
 In addition, we plot the value of $\frac{\delta \alpha
 _{\rm{min}}}{2}e^{-\frac{1}{2}}\frac{1}{4\gamma _{\rm{max}}} $ with a
 horizontal dashed line, where $\delta \alpha
 _{\rm{min}} =\min_j[\delta \alpha _j]$ and $\gamma _{\rm{max}}=\max
 _j[\gamma _j]$, and plot a vertical line at $t=\frac{1}{4\gamma _{\rm{max}}}$. We numerically show
 that we can satisfy $\tau _j(t_j)=\frac{\delta \alpha
 _{\rm{min}}}{2}e^{-\frac{1}{2}}\frac{1}{4\gamma _{\rm{max}}}$ for all $j$ by choosing a certain set
 of $\{t_j\}_{j=1}^{200}$ for $t_j\leq \frac{1}{4\gamma _{\rm{max}}} $.
 }
 \label{dcrandom}
\end{center}
\end{figure}

However, in actual experiments,  $\alpha ^{(k)}_0$, $\alpha
^{(k)}_1$, and $\gamma _k$ have a dependency on $k$ due to inhomogeneities.
In this case, we need to choose a suitable set of $t_k$ $(k=1,2,3,4)$ to
compensate for such an inhomogeneity.
%From the calculation in the Eq. \ref{ninhomodc}, we know that,
If $\tau _k(t_k)\equiv \frac{\alpha _0^{(k)}-\alpha _1^{(k)}}{2}e^{-2\gamma
 _{k}t_{k}}t_{k}$ does not depend on $k$, we can measure $B_x$ from
 $\langle\hat{N}_x^{\rm{(total)}}\rangle$ as described in Eq. (\ref{ninhomodc}).
 We numerically checked that it is possible to have an equal value of $\frac{\alpha
_0^{(k)}-\alpha _1^{(k)}}{2}e^{-2\gamma _{k}t_{k}}t_{k}$ for $k=1,2,3,4$.
In Fig. \ref{dcrandom}, we
randomly picked up $\delta \alpha _k=\alpha
_0^{(j)}-\alpha _1^{(j)}$ and $\gamma _j$ from the Gaussian
distribution, and we plotted $\tau _j(t)=\frac{\alpha _0^{(j)}-\alpha
_1^{(j)}}{2}e^{-2\gamma _{j}t}t$ ($j=1,2,\cdots ,200$). In addition, in
the same figure, we plotted the value of
$\frac{\delta \alpha
 _{\rm{min}}}{2}e^{-\frac{1}{2}}\frac{1}{4\gamma _{\rm{max}}} $ with a
 dashed line where $\delta \alpha
 _{\rm{min}} =\min_j[\delta \alpha _j]$ and $\gamma _{\rm{max}}=\max
 _j[\gamma _j]$.
% $\min_k[\max_{t}\frac{\alpha _0^{(k)}-\alpha
% _1^{(k)}}{2}e^{-2\gamma _{k}t}t ]=\min_k[\max_{t}\frac{\alpha _0^{(k)}-\alpha
% _1^{(k)}}{8\gamma _k}e^{-\frac{1}{2}} ]$.
These results show that we can choose $t_j$ to satisfy $\frac{\alpha _0^{(j)}-\alpha
_1^{(j)}}{2}e^{-2\gamma _{j}t_{j}}t_{j}=\frac{\delta \alpha
 _{\rm{min}}}{2}e^{-\frac{1}{2}}\frac{1}{4\gamma _{\rm{max}}}$ and
 $t_j\leq \frac{1}{4\gamma _{\rm{max}}}$
as long as the inhomogeneous width of
the parameters is approximately
$10\%$ \cite{acosta2009diamonds}.
%$\frac{\delta \alpha
%_{\rm{min}}}{2}e^{-\frac{1}{2}}\frac{1}{4\gamma _{\rm{max}}}$
% where $\delta \alpha
% _{\rm{min}} ={\rm{Min}}_k[\alpha
% _0^{(k)}-\alpha _1^{(k)}]$ and $\gamma
% _{\rm{max}}={\rm{Max}}_{k}[\gamma _{k}]$.
The expected values of the emitted photons from this
state are described as
\begin{eqnarray}
 \langle\hat{N}_x^{\rm{(total)}}\rangle
 &\simeq &
    \sum_{k=1}^{4}\frac{\alpha _0^{(k)}+\alpha
  _1^{(k)}}{2}\nonumber \\
 &-&\sum_{k=1,4}\frac{\alpha _0^{(k)}-\alpha _1^{(k)}}{2} g\mu
   _b\vec{B}\cdot \vec{d}_{k}t_{k}e^{-2\gamma _{k}t_{k}}\nonumber \\
  &+&\sum_{k=2,3}\frac{\alpha _0^{(k)}-\alpha _1^{(k)}}{2} g\mu
   _b\vec{B}\cdot \vec{d}_{k}t_{k}e^{-2\gamma _{k}t_{k}}\nonumber \\
     &=&(\sum_{k=1}^{4}\frac{\alpha _0^{(k)}+\alpha
  _1^{(k)}}{2})
   +\frac{2e^{-\frac{1}{2}}}{\sqrt{3}}\frac{\delta \alpha _{\rm{min}}}{4\gamma _{\rm{max}}} g\mu _bB_x\nonumber 
\end{eqnarray}
where $\delta \alpha _{\rm{min}} =\min_{k=1,2,3,4}[\alpha
_0^{(k)}-\alpha _1^{(k)}]$ and $\gamma
_{\rm{max}}=\max_{k=1,2,3,4}[\gamma _k]$.

\begin{figure}[h] 
\begin{center}
\includegraphics[width=0.9\linewidth]{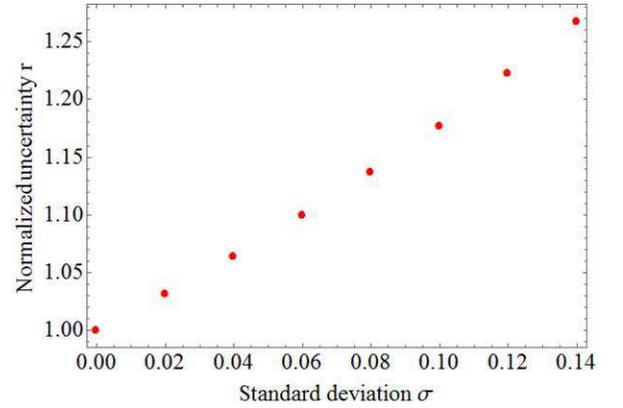} 
\caption{The normalized uncertainty of the estimation $r=\delta B^{(\rm{DC})}_x (\sigma
 )/\delta B^{(\rm{DC})}_x(\sigma =0)=\delta B^{(\rm{AC})}_x (\sigma
 )/\delta B^{(\rm{AC})}_x(\sigma =0)$ where $\delta B^{(\rm{DC})}_x (\sigma
 )$ ($\delta B^{(\rm{AC})}_x (\sigma
 )$) denotes the uncertainty in our DC (AC)
 vector magnetic field sensor for inhomogeneous parameters with a
 standard deviation of $\sigma $.
 % To calculate the average values,
 Note that the normalized uncertainty for the DC
 sensing has the same form as that for the AC sensing.
 To calculate the average value, we randomly pick up the values of $\delta \alpha _j=\alpha
_0^{(j)}-\alpha _1^{(j)}$ and $\gamma _j$ from the Gaussian
distribution where
 the average of $\delta \alpha _j$ ($\gamma '_j$) is $\overline{\delta \alpha _j}=0.01$ ($\overline{\gamma '_j}=10^6$ Hz) and the
 standard deviation is $\overline{\delta \alpha _j}\cdot \sigma '$
 ($\overline{\gamma '_j}\cdot \sigma '$) where $\sigma '$ denotes a
 normalized standard deviation.
 }
 \label{dcscale}
\end{center}
\end{figure}

Therefore, the uncertainty in the estimation of $B_x$ is given as
follows.
\begin{eqnarray}
 \delta B^{(\rm{DC})}_x&=& \frac{\sqrt{\langle \delta \hat{N}_x^{\rm{(total)}}  \delta
  \hat{N}_x^{\rm{(total)}}\rangle }}{|\frac{d\langle \hat{N}_x^{\rm{(total)}}\rangle
  }{dB_x}|}\frac{1}{\sqrt{N}}\nonumber \\
 &\simeq &\frac{\sqrt{3}\sqrt{    \sum_{k=1}^{4}\frac{\alpha _0^{(k)}+\alpha
  _1^{(k)}}{2}}}{2e^{-\frac{1}{2}}\delta \alpha _{\rm{min}}g\mu _b\sqrt{\frac{1}{4\gamma _{\rm{max}}}}}\frac{1}{\sqrt{T}}\label{ourinhomo}
\end{eqnarray}
% Therefore, as long as the inhomogeneity of the parameters is small, 
% sensitivity of our scheme is approximately 4 times better than that in the
% conventional scheme by comparing the Eq. \ref{ourinhomo} with the
% Eq. \ref{conventional}.
We numerically calculated this sensitivity, and plotted the ratio
 between the homogeneous case and inhomogeneous case with a standard deviation of
 $\sigma $
 as shown in Fig. \ref{dcscale}.
 These results demonstrate that, if the standard deviation of the parameters is around a few $\%$, we can
 achieve nearly the same sensitivity as that in the homogeneous case.

\section{AC vector magnetic field sensor via the multi-frequency control}
%We also explain how to measure $B^{(\rm{AC})}_x$ in our scheme.
Here, we explain how to measure the AC vector magnetic field using our scheme.
%component along $[1,0,0]$ in our scheme.
As an example, we discuss the case of measuring the x-component of the
AC magnetic fields.
We use a similar pulse sequence as that in our DC magnetic field sensing.
  The only difference from the DC magnetic field sensing is that we apply
a $\pi$ pulse in the middle of the microwave pulse sequence
as
shown in  the Fig. \ref{ourpulse} (b).
After the green laser irradiation, the state of the photons can be
described as follows
\begin{eqnarray}
 \rho _{\rm{AC}}^{\rm{(ph)}}= \bigotimes _{k=1}^4\rho _{k,\rm{AC}}^{\rm{(ph)}}
\end{eqnarray}
where
\begin{eqnarray}
 &&\rho ^{(\rm{ph})}_{k,\rm{AC}}=\frac{1+e^{-2\gamma '_kt_k}\sin
  \theta_k^{\rm{(AC)}}}{2} \rho ^{(\rm{ph})}_{k,0}\nonumber \\
  &+&\frac{1-e^{-2\gamma '_kt_k}\sin \theta_k^{\rm{(AC)}}}{2} \rho ^{(\rm{ph})}_{k,1}\nonumber
\end{eqnarray}
for $k=2,3$ and
\begin{eqnarray}
 &&\rho ^{(\rm{ph})}_{k,\rm{AC}}=\frac{1-e^{-2\gamma '_{k}t_{k}}\sin
  \theta_k^{\rm{(AC)}}}{2} \rho ^{(\rm{ph})}_{k,0}\nonumber \\
  &+&\frac{1+e^{-2\gamma '_{k}t_{k}}\sin \theta_{k}^{\rm{(AC)}}}{2} \rho
  ^{(\rm{ph})}_{k,1}\nonumber 
\end{eqnarray}
for $k=1,4$.
We can calculate the expected values of the emitted photon from these
state as follows
\begin{eqnarray}
 \langle\hat{N}_x^{\rm{(total)}}\rangle 
 \simeq (\sum_{k=1}^{4}\frac{\alpha _0^{(k)}+\alpha
  _1^{(k)}}{2})\ \ \ \ \ \ \ \ \ \ \ \ \ \ \ \ \ \ \ \ \ \ \ \ \ \ \ \ \
  \ \ \ \ \ \ \ \ \ \ \ \ \ \ \ \ \ \ \ 
  \nonumber \\
  -
  \sum_{k=1,4}\frac{(\alpha _0^{(k)}-\alpha
  _1^{(k)})g\mu_b\vec{B}\cdot \vec{d}_k\frac{1+\cos \omega
  _{\rm{AC}}t_k-2\cos \frac{\omega _{\rm{AC}}t_k}{2}}{\omega
  _{\rm{AC}}}e^{-2\gamma '_kt_k}}{2}
  \nonumber \\
 +\sum_{k=2,3}\frac{(\alpha _0^{(k)}-\alpha
  _1^{(k)})g\mu_b\vec{B}\cdot \vec{d}_k\frac{1+\cos
  \omega _{\rm{AC}}t_k-2\cos \frac{\omega _{\rm{AC}}t_k}{2}}{\omega
  _{\rm{AC}}}e^{-2\gamma '_kt_k}}{2}.\nonumber 
\end{eqnarray}
If we have $\alpha ^{(k)}_0=\alpha _0$, $\alpha
^{(k)}_1=\alpha _1$, $\gamma _k=\gamma $, and $t_k=t$ for all $k$, we obtain
\begin{eqnarray}
&&\langle\hat{N}_x^{\rm{(total)}}\rangle  \simeq
 2(\alpha _0+\alpha
  _1)\nonumber \\
  &-&
 %\frac{2}{\sqrt{3}}
 \frac{2(\alpha _0-\alpha
  _1)g\mu_bB_x(1+\cos \omega _{\rm{AC}}t-2\cos
  \frac{\omega _{\rm{AC}}t}{2})e^{-2\gamma 't}}{\sqrt{3}\ \omega _{\rm{AC}}}\nonumber \\
\end{eqnarray}
Note that this expectation value only depends on $B_x$.
%where $\hat{N}_k= \sum_{n=0}^{\infty }|n\rangle _{{\rm{ph}},k}\langle
%n|$
Therefore, the uncertainty in the estimation of $B_x$ is given as
follows.
\begin{eqnarray}
 \delta B^{(\rm{AC})}_x&=& \frac{\sqrt{\langle \delta \hat{N}_x^{\rm{(total)}}  \delta
  \hat{N}_x^{\rm{(total)}}\rangle }}{|\frac{d\langle \hat{N}_x^{\rm{(total)}}\rangle
  }{dB_x}|}\frac{1}{\sqrt{N}}\nonumber \\
 &\simeq &\frac{\sqrt{2(\alpha _0+\alpha _1)}}{ \frac{2}{\sqrt{3}} |\alpha _0-\alpha
  _1|g\mu_b\frac{|1+\cos \omega _{\rm{AC}}t-2\cos \frac{\omega _{\rm{AC}}t}{2}|}{\omega _{\rm{AC}}}e^{-2\gamma 't}}\frac{1}{\sqrt{\frac{T}{t}}}\nonumber \\
  \label{ouractwo}
\end{eqnarray}
where $N=\frac{T}{t}$ denotes the repetition number of the experiment.
By optimizing the parameters, we obtain
\begin{eqnarray}
 \delta B^{(\rm{AC})}_x
 &\simeq &\frac{\sqrt{3}\sqrt{2(\alpha _0+\alpha _1)}}{ 2e^{-\frac{1}{2}} |\alpha _0-\alpha
  _1|g\mu_b\frac{|1+\cos \theta _{\rm{opt}}-2\cos \frac{\theta
  _{\rm{opt}}}{2}|}{\theta _{\rm{opt}}}\sqrt{\frac{1}{4\gamma '}}}\frac{1}{\sqrt{T}}\nonumber
  \label{ourac}
\end{eqnarray}
Because  $\alpha _0\simeq \alpha _1$ , we can conclude that the
sensitivity of our scheme is approximately four times better than that in the
conventional scheme by comparing  Eq. (\ref{ourac}) with the
Eq. (\ref{conventionalac}).

  \begin{figure}[h] 
\begin{center}
\includegraphics[width=0.99\linewidth]{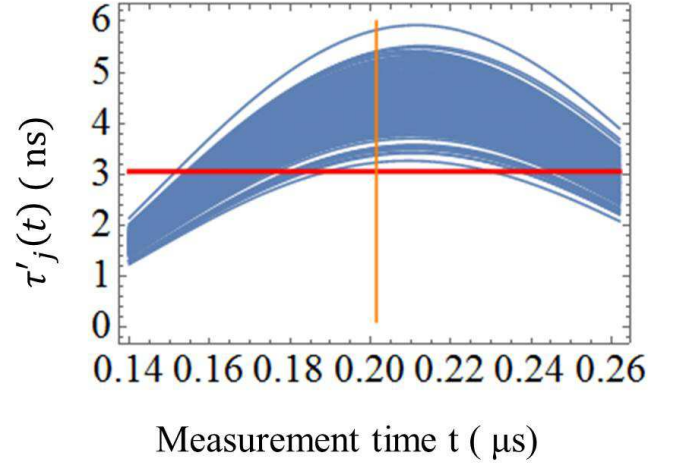} 
\caption{We plot $\tau '_j(t)=\frac{\alpha _0^{(j)}-\alpha
_1^{(j)}}{2}e^{-2\gamma '_{j}t}\frac{1+\cos (\omega _{\rm{AC}}t)-2\cos
 (\frac{\omega _{\rm{AC}}}{2}t)}{\omega _{\rm{AC}}}$
 ($j=1,2,\cdots ,200$) against $t$ where
 we choose $\delta \alpha _j=\alpha
_0^{(j)}-\alpha _1^{(j)}$ and $\gamma '_j$ from the Gaussian
distribution. 
 The average of $\delta \alpha _j$ ($\gamma '_j$) is $0.01$ ($10^6$), and the
 standard deviation is $0.001$ ($10^5$).
 In addition, we plot the value of $\frac{\delta \alpha
 _{\rm{min}}}{2}e^{-\frac{1}{2}}\frac{1+\cos (\theta _{\rm{opt}})-2\cos
 (\frac{\theta _{\rm{opt}}}{2})}{\omega _{\rm{AC}}} $ with a
 horizontal dashed line, and we plot a vertical line at $t=\frac{1}{4\gamma
' _{\rm{max}}}$ where $\delta \alpha
 _{\rm{min}} =\min_k[\delta \alpha _j]$, $\gamma '_{\rm{max}}=\max
 _j[\gamma '_j]$, and $\omega _{\rm{AC}}=4\theta _{\rm{opt}}\gamma '_{\rm{max}}$. We numerically show
 that we can satisfy $\tau '_j(t_j)=\frac{\delta \alpha
 _{\rm{min}}}{2}e^{-\frac{1}{2}}\frac{1+\cos (\theta _{\rm{opt}})-2\cos
 (\frac{\theta _{\rm{opt}}}{2})}{\omega _{\rm{AC}}}$ for all $j$ by choosing a certain set
 of $\{t_j\}_{j=1}^{200}$ where $t_j\leq \frac{1}{4\gamma '_{\rm{max}}} $.
 }
 \label{acrandom}
\end{center}
\end{figure}

Conversely,
if the parameters $\alpha ^{(k)}_0$, $\alpha
^{(k)}_1$, and $\gamma '_k$ have a dependency on $k$,
we need to choose a suitable set of $t_k$ $(k=1,2,3,4)$ to
compensate such an inhomogeneity.
We know that,
if
$(\alpha _0^{(k)}-\alpha
  _1^{(k)})e^{-2\gamma '_kt_k}\frac{1+\cos
  \omega _{\rm{AC}}t_k-2\cos \frac{\omega _{\rm{AC}}t_k}{2}}{\omega
  _{\rm{AC}}}$ does not depend on $k$, we can estimate the value of
  $B_x$ from just
  $\langle\hat{N}_x^{\rm{(total)}}\rangle$.
  We numerically checked if it is possible to have an equal value of $(\alpha _0^{(k)}-\alpha
  _1^{(k)})e^{-2\gamma '_kt_k}\frac{1+\cos
  \omega _{\rm{AC}}t_k-2\cos \frac{\omega _{\rm{AC}}t_k}{2}}{\omega
  _{\rm{AC}}}$.
In Fig. \ref{acrandom}, we
randomly picked $\delta \alpha _j=\alpha
_0^{(j)}-\alpha _1^{(j)}$ and $\gamma '_j$ from the Gaussian
distribution. We plotted $\tau '_j(t)=\frac{\alpha _0^{(j)}-\alpha
_1^{(j)}}{2}e^{-2\gamma '_{j}t}\frac{1+\cos (\omega _{\rm{AC}}t)-2\cos (\frac{\omega _{\rm{AC}}}{2}t)}{\omega _{\rm{AC}}}$
 ($j=1,2,\cdots ,200$) and the value of
$\frac{\delta \alpha
 _{\rm{min}}}{2}e^{-\frac{1}{2}}\frac{1+\cos (\theta _{\rm{opt}})-2\cos
 (\frac{\theta _{\rm{opt}}}{2})}{\omega _{\rm{AC}}}$ where $\delta \alpha
 _{\rm{min}} =\min_j[\delta \alpha _j]$, $\gamma '_{\rm{max}}=\max
 _j[\gamma '_j]$, and $\omega _{\rm{AC}}=4\theta _{\rm{opt}}\gamma '_{\rm{max}}$.
These results show that we can choose $t_j$ to satisfy
$\tau '_j(t_j)=\frac{\delta \alpha
 _{\rm{min}}}{2}e^{-\frac{1}{2}}\frac{1+\cos (\theta _{\rm{opt}})-2\cos
 (\frac{\theta _{\rm{opt}}}{2})}{\omega _{\rm{AC}}}$ for all $j$ where $t_j\leq \frac{1}{4\gamma '_{\rm{max}}} $.

 We can calculate the expected values of the emitted photons from this
state as follows:
\begin{eqnarray}
 &&\langle\hat{N}_x^{\rm{(total)}}\rangle \nonumber \\
  &\simeq &(\sum_{k=1}^{4}\frac{\alpha _0^{(k)}+\alpha
  _1^{(k)}}{2})\nonumber \\
  &-&
  \frac{2e^{-\frac{1}{2}}\delta \alpha _{\rm{min}}g\mu_bB_x(1+\cos \theta
  _{\rm{opt}}-2\cos \frac{\theta _{\rm{opt}}}{2})}{\sqrt{3}\omega _{\rm{AC}}}\nonumber \\
 \end{eqnarray}
where $\delta \alpha
 _{\rm{min}} =\min_{k=1,2,3,4}[\delta \alpha _k]$, $\gamma '_{\rm{max}}=\max
 _{k=1,2,3,4}[\gamma '_k]$, and $\omega _{\rm{AC}}=4\theta _{\rm{opt}}\gamma '_{\rm{max}}$.
Therefore, the uncertainty is
\begin{eqnarray}
 \delta B^{(\rm{AC})}_x
 &\simeq &\frac{\sqrt{3}\sqrt{\sum_{k=1}^{4}\frac{\alpha _0^{(k)}+\alpha
  _1^{(k)}}{2}}}{ 2e^{-\frac{1}{2}} \delta \alpha _{\rm{min}}g\mu_b\frac{|1+\cos \theta _{\rm{opt}}-2\cos \frac{\theta
  _{\rm{opt}}}{2}|}{\theta _{\rm{opt}}}\sqrt{\frac{1}{4\gamma
  '_{\rm{max}}}}}\frac{1}{\sqrt{T}}\nonumber \\
  \label{ourinhomoac}
\end{eqnarray}
% Therefore, as long as the inhomogeneity of the parameters is small, 
% sensitivity of our scheme is approximately 4 times better than that in the
% conventional scheme by comparing the Eq. \ref{ourinhomoac} with the
% Eq. \ref{conventionalac}.
Similar to the case of DC vector magnetic field sensing, we can 
 %These results show that,
 achieve nearly the same sensitivity as that in the homogeneous case if
 the standard deviation of the parameters is around a few $\%$ as shown in Fig. \ref{dcscale}.

In conclusion,  we proposed a scheme to improve the sensitivity of the vector
 magnetic field sensing
via multi-frequency control. 
Implementing a Ramsey interference or
 spin echo experiment for all NV centers with different NV axes using
 frequency selectivity,
we can
 enhance the signal from the NV centers. We demonstrated that the
 sensitivity of the vector magnetic field sensing becomes approximately four times better
 than that of the conventional scheme.

We thank Suguru Endo for useful discussion.
This work was supported by JSPS KAKENHI Grant No.
15K17732. This work was also supported by MEXT
KAKENHI Grants No. 15H05868, No. 15H05870, No. 15H03996, No. 26220602 and No. 26249108.
This work was also supported by Advanced Photon Science Alliance (APSA), JSPS core-to-core Program and Spin-NRJ.
%\bibliography{6mylibrary}

\end{document}